\documentclass[fleqn,12pt,twoside]{article}
\usepackage{espcrc1}
\usepackage{epsfig}       
\usepackage{amssymb}  

\newcommand{\AmS}{{\protect\the\textfont2
  A\kern-.1667em\lower.5ex\hbox{M}\kern-.125emS}}

\newcommand{\bp}{{\bf p}}
\newcommand{\bq}{{\bf q}}
\newcommand{\bk}{{\bf k}}

\newcommand{\beq}{\begin{equation}}
\newcommand{\eeq}{\end{equation}}
\newcommand{\bea}{\begin{eqnarray}}
\newcommand{\eea}{\end{eqnarray}}

\newcommand{\rmd}{{\rm d}}
\newcommand{\nn}{\nonumber}

\newcommand{\bean}{\begin{eqnarray*}}
\newcommand{\eean}{\end{eqnarray*}}

\title{\vspace*{-0.3cm}
Nonequilibrium Quantum Fields and the Classical Field Theory 
Limit
}

\author{J.\ Berges\address[]{Institute for Theoretical Physics,
        Philosophenweg 16, 69120 Heidelberg, Germany}}
       
\begin{document}

\maketitle

\begin{abstract}
\noindent
We calculate the far-from-equilibrium dynamics and 
thermalization both for the quantum and the classical $O(N)$--model.
The early and late-time behavior can be described from 
the $2PI$--loop expansion for weak couplings or the 
nonperturbative $2PI$--$1/N$ expansion 
of the effective action beyond leading order. A comparison with exact
simulations in $1+1$ dimensions in the classical limit
shows that the $2PI$--$1/N$ expansion at next-to-leading order
gives quantitatively precise results
already for moderate values of $N$.  
We derive a criterion for the validity of the classical approximation
and verify it by comparing far-from-equilibrium quantum and classical 
dynamics. At late times one observes the expected
deviations due to the difference between classical and quantum thermal
equilibrium.
\end{abstract}

\vspace*{0.3cm}

In recent years we have witnessed an enormous increase of interest 
in the dynamics of quantum fields out of equilibrium. Strong motivation
comes from a wide range of applications including  
current and upcoming relativistic heavy-ion collision experiments, 
phase transitions in the early universe or the dynamics of Bose-Einstein 
condensation. Directly simulating quantum fields in real time, such
as solving the Schr{\"o}dinger equation for the wave functional 
is prohibitively difficult and one has to find suitable approximations.
The nonequilibrium time evolution is inherently nonperturbative
in the sense that approximations based on a finite order in 
standard perturbation theory break down at sufficiently late times. 
Practicable approximations for nonequilibrium dynamics may be based on the
two-particle irreducible ($2PI$) generating functional for Green's functions 
\cite{Cornwall:1974vz,Calzetta:1988cq}. Recently, the $2PI$ effective
action has been solved for a $1\!+\!1$ dimensional scalar quantum field theory 
at next-to-leading order in the $2PI$--loop expansion 
\cite{Berges:2000ur} and in the $2PI$--$1/N$ expansion \cite{Berges:2001fi}.
Both the far-from-equilibrium early-time behavior and the late-time physics 
of thermalization were successfully described. 
For a recent review on the use of the $2PI$ effective action
in nonequilibrium field theory see Ref.\ \cite{review}.

A unique possibility to calculate the {\em exact} time evolution, 
which includes all orders in loops or $1/N$, is provided by the
classical statistical field theory limit. The exact evolution 
(up to statistical errors)
of correlation functions can be constructed by numerical integration
and sampling of initial conditions from a given probability 
distribution function. On the level of correlation
functions classical and quantum evolution equations are remarkably
similar\footnote{Of course, in contrast to the quantum theory the 
classical limit
suffers from Rayleigh-Jeans divergences and has to be regulated. In
$1\!+\!1$ dimensions such divergences are absent in 
$\phi\phi$-correlation functions \cite{TangSmit}.} 
and the same approximation schemes and initial conditions can be applied.  
This aspect of classical field theory has been 
stressed in Refs.\ \cite{ABW,BS,Blagoev:2001ze,AB2}. 
In Ref.\ \cite{Blagoev:2001ze}
this is applied to the next-to-leading order classical $\phi^4$-model.
In Ref.\ \cite{AB2} it is shown that the $1/N$ expansion at 
next-to-leading order
converges to the exact result by increasing $N$ already for moderate
values of $N$.  
Apart from benchmarking approximation schemes employed in quantum field
theory, the classical field limit is of great practical importance
and often applied for the approximate description of nonequilibrium 
quantum fields. In this note I want to elaborate on Ref.\ \cite{AB2},
done together with G.\ Aarts, and study the conditions under which 
far-from-equilibrium quantum dynamics can be reliably described
by classical fields. 

We consider a real \mbox{$N$-component}
scalar quantum field theory with a $\lambda
(\phi_a\phi_a)^2/(4! N)$ interaction in the symmetric
phase ($a=1,\ldots,N$). There are two linearly independent two-point 
functions which can be related to the anti-commutator and commutator of
two field operators \cite{Aarts:2001qa,Berges:2001fi}
\beq
F_{ab}(x,y)=\langle[{\phi}_a(x),{\phi}_b(y)]_+ \rangle/2 \quad ,\qquad
\rho_{ab}(x,y)=i \langle[{\phi}_a(x),{\phi}_b(y)]_- \rangle 
\eeq
Here $F$ is the ``symmetric'' propagator and $\rho$ denotes the
spectral function. The classical equivalent of the 
spectral function is obtained by replacing the commutator
by the Poisson bracket.
For the analytic presentation we consider here the three-loop expansion
of the $2PI$ effective action for $N=1$, which has been employed in Ref.\
\cite{Berges:2000ur} 
to study late-time thermalization in a quantum field theory.
Numerical results from the $1/N$ expansion of the $2PI$ effective 
action at next-to-leading order for $N>1$ are shown below.  
For spatially homogeneous fields the dynamics of the Fourier 
transformed $F$ and $\rho$ is described by \cite{Aarts:2001qa,Berges:2001fi}
\bea
\nn
\left[\partial_{t_x}^2 + M^2(t_x) \right]F(t_x,t_y;\bp)\! &=&
- \int_0^{t_x}\!\!\! \rmd t_z
\,\, \Sigma_{\rho}(t_x,t_z;\bp)F(t_z,t_y;\bp) \nonumber \\
&& + \int_0^{t_y}\!\!\! \rmd t_z \,\, \Sigma_{F}(t_x,t_z;\bp)
\rho(t_z,t_y;\bp), 
\label{eqF1}
\\ 
\left[\partial_{t_x}^2 +M^2(t_x)\right]\rho(t_x,t_y;\bp) &=&
-\int_{t_y}^{t_x}\!\!\! \rmd t_z \,\,
\Sigma_{\rho}(t_x,t_z;\bp)\rho(t_z,t_y;\bp).
\nonumber
\eea
which are exact for known $\Sigma_F$, $\Sigma_{\rho}$.
From the three-loop $2PI$ effective action the effective mass term is
$
M^2(t_x) = m^2 + (\lambda/2)
\int\! \frac{d \bq}{(2\pi)^d}\, F(t_x,t_x;\bq)
$
and the self energies are 
\bea
\Sigma_{F}(t_x,t_y;\bp) &=& -\frac{\lambda^2}{6} 
\int \frac{d \bq}{(2\pi)^d} \frac{d \bk}{(2\pi)^d}\,\,
F(t_x,t_y;\bp-\bq-\bk)\nonumber \\ && \left[F(t_x,t_y;\bq)F(t_x,t_y;\bk) 
- \frac{3}{4}\, \rho(t_x,t_y;\bq)\rho(t_x,t_y;\bk) \right], 
\nonumber\\
\Sigma_{\rho}(t_x,t_y;\bp) &=& -\frac{\lambda^2}{2} 
\int \frac{d \bq}{(2\pi)^d} \frac{d \bk}{(2\pi)^d}\,\,
\rho(t_x,t_y;\bp-\bq-\bk)
\label{sigmarho}  \\ && \left[F(t_x,t_y;\bq)F(t_x,t_y;\bk) 
- \frac{1}{12}\, \rho(t_x,t_y;\bq)\rho(t_x,t_y;\bk) \right]. 
\nonumber
\eea
The classical statistical field theory
limit of a scalar quantum field theory has been studied extensively
in the literature. An analysis along the lines of Refs.\
\cite{Aarts:1997qi,Cooper:2001bd,Blagoev:2001ze,AB2}
shows that all equations (\ref{eqF1})--(\ref{sigmarho}) remain the same in the
classical limit except for
differing expressions for the self energy:
\bea
\Sigma_{F}(t_x,t_y;\bp)\!\! & {\rm classical\, limit} \atop \Longrightarrow & 
\!\! -\frac{\lambda^2}{6} 
\int\! \frac{d \bq}{(2\pi)^d} \frac{d \bk}{(2\pi)^d}\,
F(t_x,t_y;\bp-\bq-\bk) F(t_x,t_y;\bq) F(t_x,t_y;\bk),\label{eqPicl} \\
\Sigma_{\rho}(t_x,t_y;\bp)\!\! &{\rm classical\, limit} \atop 
\Longrightarrow & 
\!\! -\frac{\lambda^2}{2} 
\int\! \frac{d \bq}{(2\pi)^d} \frac{d \bk}{(2\pi)^d}\,
\rho(t_x,t_y;\bp-\bq-\bk) F(t_x,t_y;\bq) F(t_x,t_y;\bk).
\eea
One observes that the classical self energies are obtained from the 
expressions in the quantum theory by dropping terms with two spectral
($\rho$) components compared to two statistical ($F$) functions. 
In particular, it becomes obvious that the leading order equations
(similarly for leading-order large $N$, Hartree or mean field) 
are identical for the quantum and the classical theory, and the inclusion
of direct scattering effects is crucial.

\begin{figure}[t]
\begin{center}
\epsfig{file=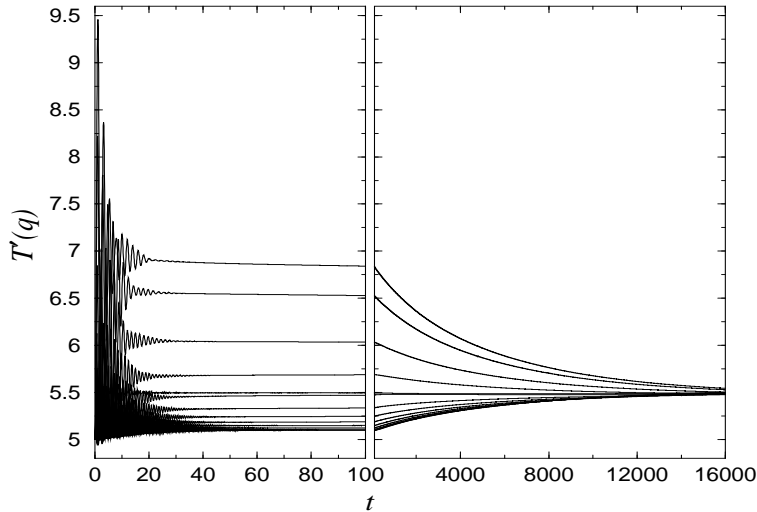,width=10.cm,height=7.cm,angle=0}
\end{center}
\vspace*{-1.5cm}
\caption{Nonequilibrium classical time evolution from the
three-loop approximation of the $2PI$ effective action for
$\lambda/m^2 =1$. Shown is the 
mode temperature $T'(t;q)$, as defined in the text, 
with $T'(t=0;q)/m = 5$ \cite{AB3}. 
The equivalent approximation for the  
corresponding quantum field theory has been used 
in Ref.\ \cite{Berges:2000ur} to demonstrate the late-time approach to quantum
thermal equilibrium. In contrast, the classical field theory  
approaches classical thermal equilibrium as one clearly observes from
the approach $T'(t;q) \to T_{\rm cl}$ which corresponds to classical 
equipartition.
\vspace*{-0.5cm}}
\label{fixedpoint}
\end{figure}

The solution of the above classical evolution equations 
is shown in Fig.\ \ref{fixedpoint} 
for Gaussian initial correlations with  
temperature $T'(t=0;q)/m = 5$ and $\lambda/m^2 =1$.\footnote{For the classical
theory we employ a lattice 
regularization with spatial lattice spacing $m a_s=0.4$
corresponding to a fixed momentum cutoff $\Lambda=\pi/a_s$.
We observe that at sufficiently late times the contributions from 
early times to the dynamics are effectively suppressed. This 
fact is shown in detail in Ref.\ \cite{AB3} and has been employed in 
Fig.\ \ref{fixedpoint} to reach the very late times.}
(For the numerical implementation 
see Ref.\ \cite{Berges:2001fi}.)
The ``mode temperature'' is defined by \cite{ABW}
\beq
T'(t_x;p) = \partial_{t_x}\partial_{t_y} 
F(t_x,t_y;p)|_{t_x=t_y} \, .
\label{modet}
\eeq
One observes that the system relaxes
to a final temperature $T_{\rm cl}/m=5.5$.
The thermalization
time turns out to be very large and we find an exponential 
late-time relaxation to thermal equilibrium with rate 
$\gamma^{\rm (therm)} \simeq 2 \times 10^{-4}$ for $T(t;q=0)$ \cite{AB3}. 
We emphasize that the exponential behavior with similarly 
long thermalization times are found as well from exact simulations 
as in Ref.\ \cite{ABW}\footnote{Note 
that these authors employ a fixed lattice cutoff $\pi/a_s=4 \pi m$.}.

\begin{figure}
\unitlength1.0cm
\begin{center}
\begin{picture}(15.,6.8)
\put(-0.7,0.){
\epsfig{file=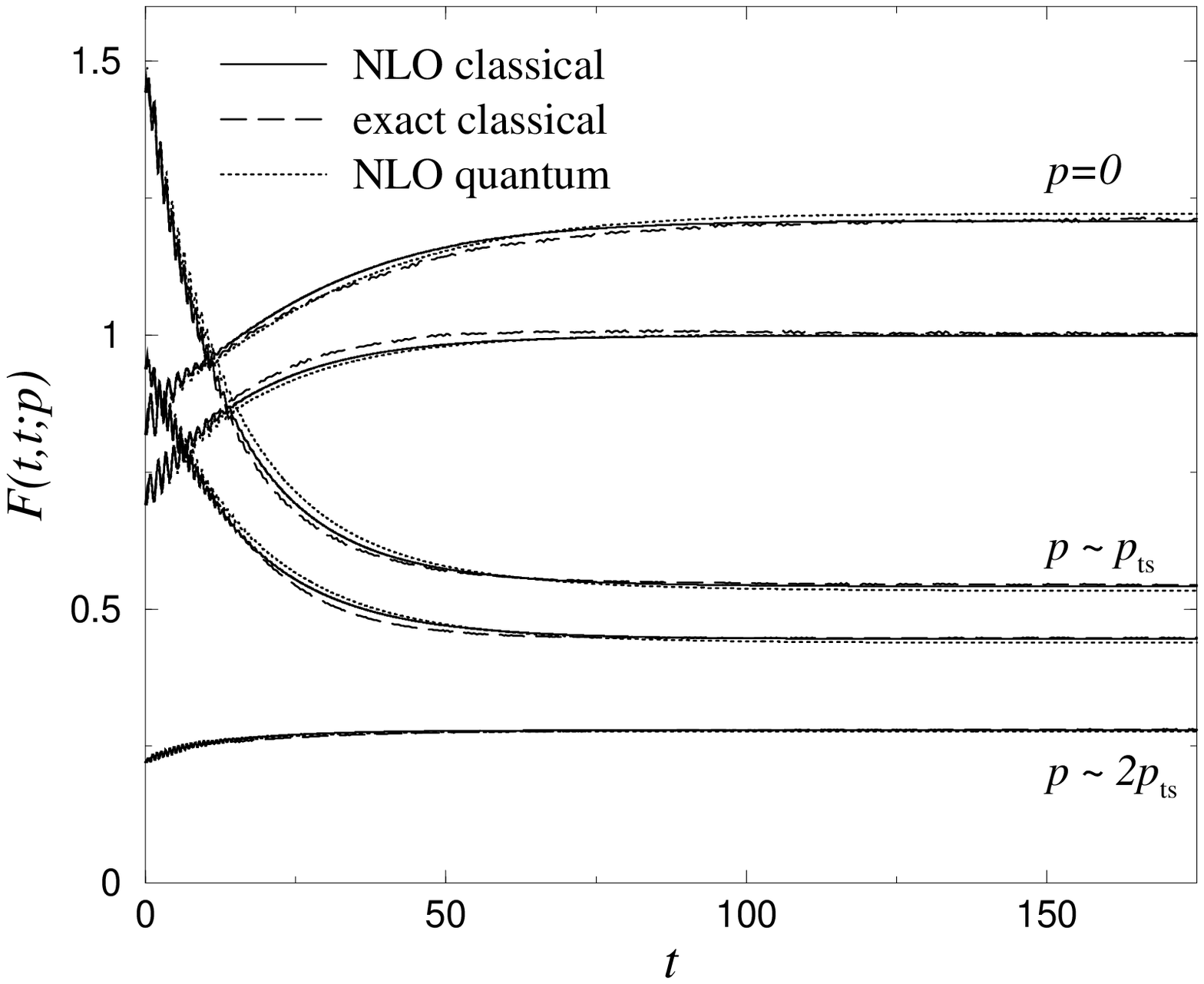,width=7.9cm,height=6.7cm,angle=0}}
\put(7.4,0.){
\epsfig{file=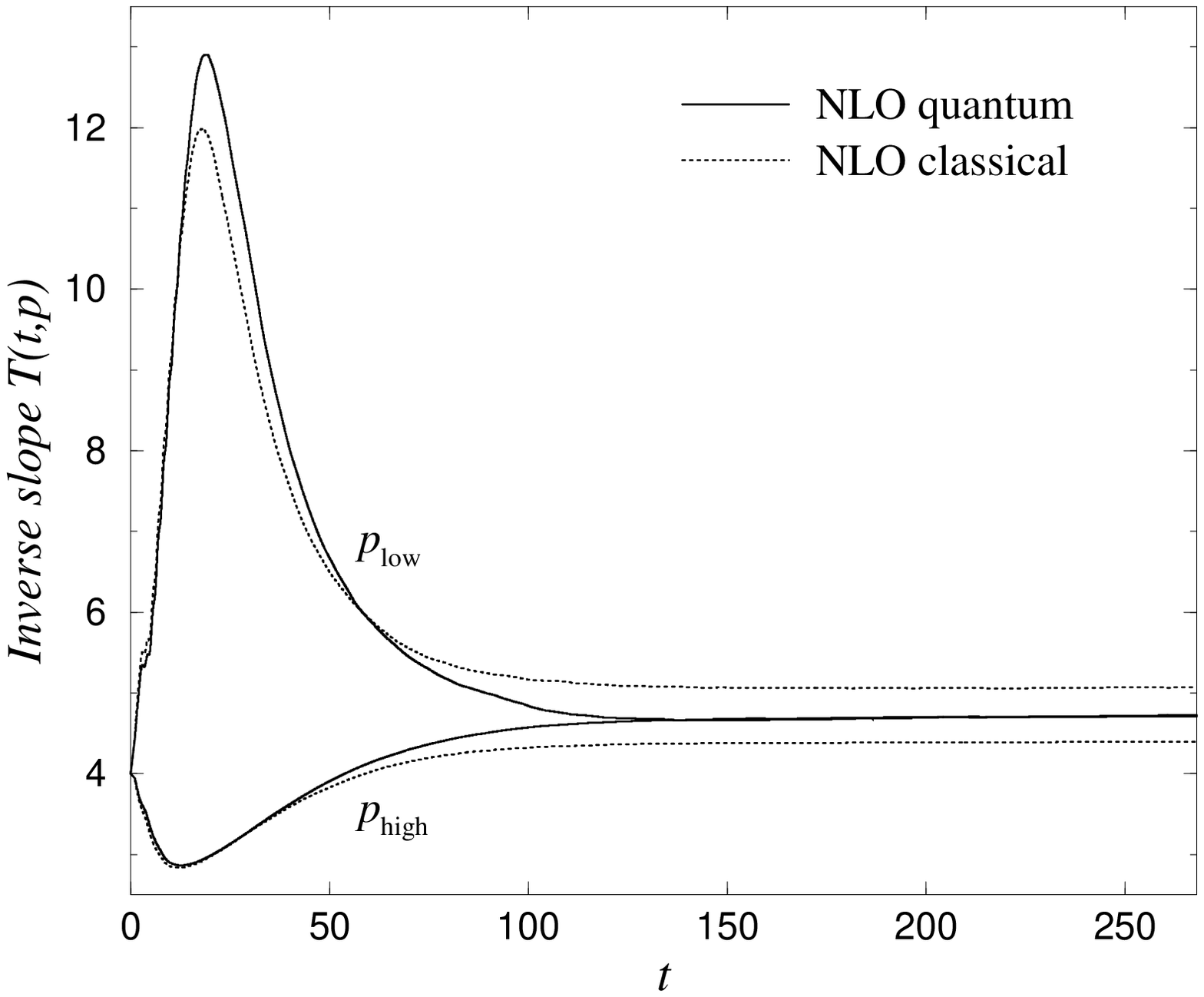,width=7.9cm,height=6.7cm,angle=0}}
\end{picture}
\end{center}
\vspace*{-1.5cm}
\caption{Left: 
far-from-equilibrium evolution of the two-point function
$F(t,t;p)$ for various momenta $p$ in units of $m_R$. 
The coupling is $\lambda/6N = 0.5 \, m_R^2$ for $N=10$. 
One observes a good agreement between the exact (dashed) and
the next-to-leading order classical result (full) \cite{AB2}. 
The quantum evolution is shown with
dotted lines for momenta $p \lesssim 2 p_{\rm ts}$, 
for which the classicality condition (\ref{classicality})
is approximately valid. Right: A very sensitive quantity to 
study deviations is the time dependent inverse slope 
$T(t,p)$ defined in the text. When quantum thermal equilibrium 
with a Bose-Einstein distributed particle number is approached all 
modes get equal $T(t,p)=T_{\rm qm}$, as can be observed to high 
accuracy for the quantum evolution \cite{Berges:2001fi,AB2}. For classical 
thermal equilibrium the defined slope remains momentum dependent
and $T'(t,p)$ becomes constant (cf.\ Eq.\ (\ref{modet})). 
\vspace*{-0.5cm}
\label{twofigs}
}
\end{figure}
The classical field approximation
is expected to become a reliable description for the quantum theory if
the number of field quanta in each mode is sufficiently high. 
Comparing Eqs.\ (\ref{sigmarho}) and (\ref{eqPicl}) one observes 
the sufficient condition for
\beq
{\rm classical\,\,\, evolution:} \quad F(t_x,t_y;\bq)F(t_x,t_y;\bk)\, \gg \,
 \frac{3}{4}\, \rho(t_x,t_y;\bq)\rho(t_x,t_y;\bk) \, .
\label{sufficient}
\eeq
To obtain an estimate on the occupation numbers we
define a time-dependent
effective particle number $n(t;p)$ and 
mode energy $\epsilon_\bp(t)$ \cite{Aarts:2001qa}
\bea
n(t_x;\bp)+\frac{1}{2}
\equiv \Big(F(t_x,t_y;\bp)\, \partial_{t_x}\partial_{t_y} 
F(t_x,t_y;\bp) \Big)^{1/2}_{|t_x=t_y}
\, , \quad 
\epsilon_\bp(t_x) \equiv \left(\, \frac{\partial_{t_x}\partial_{t_y} 
F(t_x,t_y;\bp)}{F(t_x,t_y;\bp)}\right)^{1/2}_{|t_x=t_y} \,\, .
\nonumber
\label{eqpartnr}
\eea
With these definitions $F(t_x,t_x;\bp) \equiv
(n(t_x;\bp)+1/2)/(\epsilon_\bp(t_x))$.
Note that 
$\rho \equiv 0$ for \mbox{$t_x=t_y$} and at unequal times we find
$|\rho(t_x,t_y;\bp)| \, \lesssim \,
{\rm max}[(\epsilon_\bp(t))^{-1}_{|t=[t_x,t_y]}]$.
The latter relation has indeed to be valid for the free theory or 
if the weakly coupled quasiparticle 
picture applies, which is underlying the 
above particle number and mode energy definitions. 
Time averaged over an oscillation period $2\pi/\epsilon_\bp(t)$
one can translate (\ref{sufficient}) 
into an estimate on the lower bound for the effective
particle number: 
\beq
\left[n(t;\bp)+\frac{1}{2} \right]^2 \, \gg \, \frac{3}{4}
\qquad {\rm or} \qquad n(t;\bp) \, \gg \, 0.37 \,\, .
\label{classicality}
\eeq
This limit agrees rather well with what is found in thermal equilibrium.
For a Bose-Einstein distributed particle number $n_T=(e^{\epsilon/T}-1)^{-1}$ 
with temperature $T$ one finds 
$n_T(\epsilon = T) = 0.58$, below which deviations from the 
classical thermal distribution become sizeable. 

A similar estimate can also be obtained beyond the weak coupling
regime from the $1/N$ expansion of the $2PI$ effective action at
next-to-leading order. In Fig.\ \ref{twofigs} we consider the full 
next-to-leading order time evolution in the ($2PI$--) $1/N$
expansion \cite{Berges:2001fi,AB2}. 
The far-from-equilibrium
initial particle number, $n_0(p)$, is described by a Gaussian 
$n_0(p)={\cal A} \exp ( 
-\frac{1}{2 \sigma^2}(|p|-p_{\rm ts})^2 )$
peaked around $p_{\rm ts}=2.5 m_R$ with $\sigma=0.25 m_R$ and
${\cal A}=4$. We add to $n_0$
a thermal ``background'' $n_T$ with initial temperature $T=4\, m_R$.
Here $m_R$ is the one-loop renormalized mass
in vacuum ($n \equiv 0$).

For the current initial distribution the classicality condition 
(\ref{modet}) is approximately valid for momenta $p \lesssim 2 p_{\rm ts}$
with   
$n(t=0;p=2 p_{\rm ts}) \simeq 0.35$ 
($n(t=0;p=p_{\rm ts}) \simeq 4.5$) and a slightly larger final density 
at this momentum of about $n(p=2 p_{\rm ts}) \simeq 0.5$. 
We observe a good agreement of quantum and classical evolution for this
momentum range as shown in \mbox{Fig.\ \ref{twofigs}}. The right
figure shows the time dependent inverse slope parameter \cite{Berges:2001fi}
\beq
T(t,p) \equiv - n(t,\epsilon_p) \mbox{$[n(t,\epsilon_p)+1]$} 
(dn/d\epsilon)^{-1} \, .
\eeq
This parameter is constant for a
Bose-Einstein distributed particle number and remains momentum
dependent for classical thermal equilibrium. 
In Fig.~\ref{twofigs} we plot the function
$T(t,p)$ for $p_{\rm low}\simeq 0$ and $p_{\rm high} \simeq 2 p_{\rm ts}$. 
Initially one observes a very different behavior of $T(t,p)$ for the low
and high momentum modes, indicating that the system is far from
equilibrium. The quantum evolution approaches
quantum thermal equilibrium with a momentum independent inverse slope 
$T_{\rm qm}=4.7\, m_R$ to high accuracy. 
In contrast, in the classical limit this slope parameter remains momentum
dependent and the system relaxes towards classical thermal equilibrium,
as exemplified in Fig.\ \ref{fixedpoint}. 

\vspace*{0.2cm}
I thank G.\ Aarts for collaboration on this work.


\begin{thebibliography}{9}

\bibitem{Cornwall:1974vz}
J.~M.~Cornwall, R.~Jackiw, E.~Tomboulis,
Phys.\ Rev.\ {\bf D10} (1974) 2428.     

\bibitem{Calzetta:1988cq}
E.~Calzetta, B.~L.~Hu,
Phys.\ Rev.\ {\bf D37} (1988) 2878.

\bibitem{Berges:2000ur}
J.~Berges, J.~Cox,
Phys.\ Lett.\ {\bf B517} (2001) 369 [hep-ph/0006160].

\bibitem{Berges:2001fi}
J.~Berges,
Nucl.\ Phys.\ {\bf A699} (2002) 833
[hep-ph/0105311].

\bibitem{review}
J.~Berges, hep-ph/0109170.

\bibitem{TangSmit}
W.H.\ Tang, J.\ Smit, Nucl.\ Phys.\ {\bf B540} (1999) 437.

\bibitem{ABW}
G.\ Aarts, G.F.\ Bonini, C.\ Wetterich,
Phys.\ Rev.\ {\bf D63} (2001) 025012.

\bibitem{BS}
S.\ Bors{\'a}nyi, Z.\ Sz{\'e}p, Phys.\ Lett.\ {\bf B508} (2001) 109.

\bibitem{Blagoev:2001ze}
K.~Blagoev, F.~Cooper, J.~Dawson, B.~Mihaila,
Phys.\ Rev.\ {\bf D64} (2001) 125003.

\bibitem{AB2} G.\ Aarts, J.\ Berges, Phys.\ Rev.\ Lett.\ {\bf 88} (2002) 
041603 [hep-ph/0107129].

\bibitem{Aarts:2001qa}
G.\ Aarts, J.\ Berges,
Phys.\ Rev.\ {\bf D64} (2001) 105010.

\bibitem{Aarts:1997qi}
G.\ Aarts, J.\ Smit,
Phys.\ Lett.\ {\bf B393} (1997) 395,
Nucl.\ Phys.\ {\bf B511} (1998) 451;
W.~Buchm\"uller, A.~Jakov\'ac,         
Phys.\ Lett.\  {\bf B407} (1997) 39.

\bibitem{Cooper:2001bd}
F.~Cooper, A.~Khare and H.~Rose,
Phys.\ Lett.\ {\bf B515} (2001) 463.

\bibitem{AB3} G.\ Aarts, J.\ Berges, in preparation. 

\end{thebibliography}
\end{document}